\documentclass[11pt,a4paper,iop]{emulateapj}
\usepackage{epsfig}
\usepackage{amsmath}
\usepackage{natbib}

\slugcomment{2013 April 18}

\shorttitle{}
\shortauthors{Isella A.}

\begin{document}


\title{An Azimuthal Asymmetry in the LkH$\alpha$~330 Disk}
 

\author{Andrea Isella} 
\email{isella@astro.caltech.edu}
\affil{Department of Astronomy, California Institute of Technology, MC 249-17, Pasadena, CA 91125, USA}
\author{Laura M. P\'erez}
\affil{Jansky fellow, NRAO, Socorro, NM, USA}
\author{John M. Carpenter, Luca Ricci}
\affil{Department of Astronomy, California Institute of Technology, MC 249-17, Pasadena, CA 91125, USA}
\author{Sean Andrews, Katherine Rosenfeld}
\affil{Harvard-Smithsonian Center for Astrophysics, 60 Garden Street, Cambridge, MA 02138, USA}

\begin{abstract}

Theory predicts that giant planets and low mass stellar companions shape circumstellar disks 
by opening annular gaps in the gas and dust spatial distribution. For more than a decade it has been debated whether
 this is the dominant process that leads to the formation of transitional disks.
In this paper, we present millimeter-wave interferometric observations of the transitional disk around the 
young intermediate mass star LkH$\alpha$~330. These observations reveal a lopsided ring in the 1.3 mm dust thermal 
emission characterized by a radius of about 100 AU and an azimuthal intensity variation of a factor of 2.  
By comparing the observations with a gaussian parametric model, we find that the observed asymmetry 
is consistent with a circular arc, that extends azimuthally by about 90\arcdeg\ and emits about 1/3 of the 
total continuum flux at 1.3~mm. Hydrodynamic simulations show that this structure 
is similar to the azimuthal asymmetries in the disk surface density that might be produced by the 
dynamical interaction with unseen low mass companions orbiting within 70 AU from the central star. 
We argue that such asymmetries might lead to azimuthal variations in the millimeter-wave dust opacity 
and in the dust temperature, which will also affect the millimeter-wave continuum emission.
Alternative explanations for the observed asymmetry that do not require the presence of companions 
cannot be ruled out with the existing data. Further observations of both the dust and molecular gas emission 
are required to derive firm conclusions on the origin of the asymmetry observed in the LkH$\alpha$ 330 disk.
\end{abstract}



\section{Introduction}
\label{sec:intro}

Although it is generally accepted that planets form in 
disks around young stars, the evidence supporting 
this hypothesis is circumstantial. 
Back in 1989, measurements of the spectral energy distributions (SED)
of young stellar systems revealed a family of objects 
characterized by little or none near- and mid-infrared excess, but higher luminosity 
at longer wavelengths \citep{Strom89}. It was immediately recognized that the 
lack of infrared emission might be a sign of clearing of small dust grains from the inner disk, 
as previously predicted in the 
presence of forming giant planets \citep{Lin79}. With the idea that these objects 
would be evolving to planetary systems, they were called {\it transitional disks}. 
In the last decade, infrared observations by the {\it Spitzer} space telescope revealed 
dozens of transitional disks \citep[see, e.g.,][]{Cieza12a,Currie11}, which represent 
at least 15\% of the total disk population \citep{Muzerolle10}.
However, it took almost two decades after their discovery 
to spatially resolve in the dust continuum emission, the inner cavities  predicted by SED models \citep{Pietu06}. 

Detections of companions orbiting within dust-depleted cavities 
have been obtained using aperture masking interferometric observations and speckle 
imaging at infrared wavelengths. In the case of CoKu Tau 4, the observations revealed 
that the cavity is due to a stellar mass companion orbiting at about 8 AU from the 
central star \citep{Ireland08}. This result raised the possibility that transitional disks 
might be circumbinary disks. However, observations of T Cha, LkCa~15, and TW~Hya 
suggest the presence of companions with masses below the deuterium burning limit 
\citep{Huelamo11,Kraus12,Arnold12}.  

Besides these few detections, crucial information on the nature of transitional disks 
has been obtained through spatially resolved observations of the disk emission at 
infrared and millimeter wavelengths. In particular, long baseline interferometric 
observations have revealed  cavities in the (sub-)millimeter dust continuum emission 
(called {\it millimeter cavities} henceforth) that extend out to several tens of AU from the 
central star \citep{Brown08,Brown12, Andrews09, Andrews11,Hughes09, Isella10a, 
Isella10b, Isella12, Cieza12b}. In the case of SAO~206462, LkCa~15, PDS~70, and 
2MASS J16042165--2130284, the millimeter cavities correspond to cavities in the 
near-infrared scattered light, and their sizes are in good agreement with the predictions 
of SED models \citep[][respectively]{Thalmann10,Muto12,Dong12,Mayama12}.
However, some of the most extended millimeter cavities are not accompanied by a 
deficit in the near-infrared excess \citep[][]{Pietu05,Isella10b,Andrews11} or contain a 
significant amount of molecular gas \citep{Casassus13,Dutrey08,Isella10b,Rosenfeld12}. 
Infrared and millimeter  observations also reveal that disks characterized by 
large millimeter cavities  can significantly deviate from central symmetry.
For example, the disks around the Herbig Ae stars AB~Aur, SAO~206462, 
and MWC~758, and HD~142527 show spiral arcs in the scattered light emission 
that have been interpreted as evidence of dynamical interactions by unseen 
low mass companions \citep{Hashimoto11,Muto12,Grady13}. 
In the case of AB Aur, millimeter observations reveal an asymmetric ring of dust with 
a radius of about 150 AU and spiral arms in the CO (2-1) line emission that 
correspond to spiral arcs observed in scattered light \citep{Tang12,Fukagawa04}. 
Prominent asymmetries in the millimeter emission have been also detected toward SAO~206462, 
MWC~758, and HD~142527, but their geometrical structure is not well constrained due
to the limited angular resolution of existing observations  \citep{Ohashi08,Isella10b,Andrews11,Casassus13}.
The spatially resolved observations of transitional disks obtained to date reveal 
structures that  are much more complex than those suggested by SED models but 
still lack the angular resolution and sensitivity required to explain the broad range of observed properties.

In this paper, we present new CARMA observations of the 1.3~mm dust emission toward 
LkH$\alpha$~330, a 3 Myr old, G3 pre-main sequence star in the Perseus molecular cloud at a 
distance of 250 pc \citep[M$_\star$=2.5 M$_\sun$,][]{Osterloh95,Cohen79}.
Previous SMA observations of the 0.88 mm dust emission revealed that LkH$\alpha$~330 
is surrounded by a circumstellar disk inclined by about 35\arcdeg\ with respect to the line 
of sight that extends out to at least 130 AU from the central star. These observations also 
revealed a millimeter cavity  with a radius of 40 AU and a possible azimuthal 
asymmetry in the dust emission outside the cavity \citep{Brown08}.  
Near-infrared Keck aperture masking observations exclude the presence of stellar mass 
companions at a projected separation larger than 40 mas, which corresponds to orbital 
radii between 10 AU and 13 AU at the distance of LkH$\alpha$~330 and for a disk inclination of 35\arcdeg\
(A. Kraus private communication). These observations leave open the possibility that 
the large millimeter cavity might be cleared by brown dwarfs or planetary size companions. 
More recent SMA observations revealed that the millimeter cavity is larger, about 70 AU in radius, 
than estimated by \cite{Brown08},  and that the overall dust emission can be explained reasonably 
well with a symmetric disk model \cite[][]{Andrews11}. The CARMA observations presented in this paper 
achieve better image fidelity than previous SMA data, which allows us 
to derive firm conclusions on the presence of asymmetries in LkH$\alpha$~330 disk.

The paper is organized as follows. The 
CARMA obserations are presented in Section~\ref{sec:obs}. In Section~\ref{sec:ana} we analyse the
CARMA data, as well as the previously published SMA data, to constrain the geometrical properties 
of the asymmetries observed in the dust emission. In Section~\ref{sec:disc} we discuss the physical 
processes that can lead to the formation of azimuthal asymmetries in the dust emission. We
present our conclusions in Section~\ref{sec:conc}.

\section{Observations}
\label{sec:obs}

LkH$\alpha$~330 was observed between December 2008 
and January 2009 using the CARMA A and B array configurations, 
which provide baseline lengths in the range 82-1900~m. Receivers
were tuned at the frequency of 230 GHz ($\lambda = 1.3$~mm) and 
the correlator was configured to use a total bandwidth of 4 GHz to 
maximize the sensitivity to the continuum.  Atmospheric conditions
during the observations  were excellent, with zenith opacity 
$\tau_{230\textrm{GHz}} < 0.15$ and rms noise phase coherence 
below 100 $\mu$m as measured on a 100 m baseline.
 
Data were calibrated using the Multichannel Image Reconstruction, Image Analysis and
Display (MIRIAD) software package (Sault et al. 1995).
The radio galaxy 3C111, separated by 8.7\arcdeg\ from LkH$\alpha$~330, 
was observed every 7 minutes to measure the complex antenna 
gains and system bandpass shape. Further, on the longest baselines, the CARMA Paired 
Antenna Calibration System \citep[C-PACS,][]{Perez10}, was applied to derive antenna 
gains on a time scale of 10 seconds. By simultaneous observations of a nearby point 
source we estimate that the phase decoherence after antenna gain calibration 
corresponds to a seeing of less than 0.05\arcsec. Absolute flux density calibration was obtained 
by observing Uranus, and by comparing the flux density of 3C111 with almost simultaneous 
SMA observations; the flux density uncertainty is estimated to be 10\%.

We also analyze SMA observations of the 0.88~mm dust continuum emission 
measured toward LkH$\alpha$~330 previously published by \cite{Brown08} and \cite{Andrews11}. 
The details of the adopted receiver and  the correlator configuration are described in these 
two works and will not be repeated here. The SMA observations were obtained with 
the ``very extended"  array configuration on November 2006 and with the 
compact configuration on November 2010. Once combined, the data provide 
baseline lengths between 8~m and 509~m.  The SMA data were calibrated 
using the MIR software package. The bandpass response of the system was set by observing 
the nearby quasars 3C273, 3C454.3, and 3C84, while the complex 
antenna-based gain were calibrated by observing 3C111. 
The absolute flux scale of the visibilities was derived by observing Uranus and Titan, 
resulting in a systematic uncertainty of about 10\%.
 
To properly combine observations taken at different epochs, 
we shifted the phase center of each track by the known stellar proper motion.
We adopted the J2000 coordinates and proper motion from 
the UCAC4 catalog  \citep{Zacharias13}: RA=03h~45m~48.282s~$\pm$~0.001s, 
Dec=+32\arcdeg\ 24\arcmin\ 11.85\arcsec\ $\pm$ 0.02\arcsec\,
pmRA=$3.9\pm2.0$~mas/yr, and pmDec=$-6.5\pm2.1$~mas/yr.
The SMA tracks obtained in the "very extended" and compact array configurations 
were corrected by assuming a time baseline of 6.8~yr and 10.8~yr, respectively. 
The time baselines for the CARMA A and B array configuration observations are 
8.9~yr and 9.0~yr, respectively. The same time baselines were used to derive the 
uncertainty on the modern position of the star due both to the errors 
on the J2000 coordinates and on the proper motion. We calculate that 
the absolute uncertainty on the stellar position at the time of the observations 
varies between  35~mas and 40~mas. 

Errors in the centering position of each track might in principle produce spurious 
asymmetric structures in the image resulting from their combination. However, 
since the CARMA observations were obtained only one month apart, 
the error in the relative centering position due to the stellar proper motion is 
less than  one milli-arcsecond. In the case of SMA data, the observations are
separated by four years, which lead to a possible centering error of about 8
mas. In both cases, the centering errors should be much smaller than the 
angular resolution of the observations and should not significantly affect the 
morphology of the dust continuum emission.

\section{Morphology of the dust emission}
\label{sec:ana}

\subsection{Image domain}
\begin{figure}[!t]
\centering
\includegraphics[angle=0, width=\linewidth]{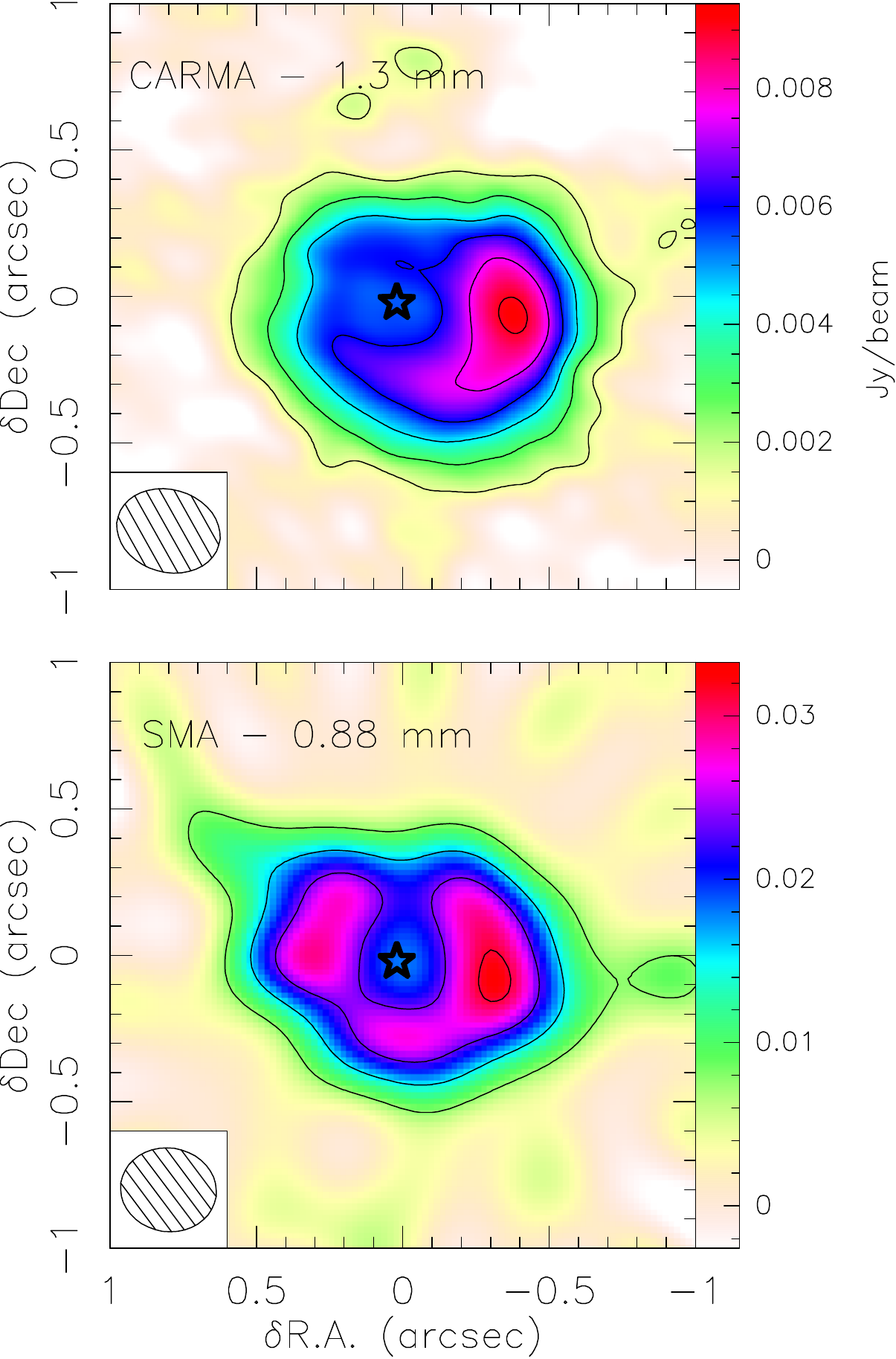}
\caption{\label{fig:obs} Maps of the continuum emission observed toward LkH$\alpha$~330 at the 
wavelength of 1.3 mm (top) and 0.88 mm (bottom). The two maps have a similar angular resolution of 
about 0.35\arcsec. The intensity contours are spaced by 3 times the noise level, where the noise
level is to 0.5 mJy/beam and 2.6 mJy/beam in the top and bottom panel respectively. The position 
of the central star is indicated with the star symbol.}
\end{figure}

Figure~\ref{fig:obs} presents the images of the LkH$\alpha$~330 disk 
at the wavelengths of 1.3~mm and 0.88~mm as obtained with CARMA and 
SMA, respectively.
The CARMA observations achieve an angular resolution of 0.35\arcsec, 
or a spatial resolution of 90 AU at the distance of the source, as 
obtained by applying natural weighting to the {\it uv} data. 
The SMA image has similar resolution with uniform weighting, which boosts
the angular resolution by down-weighting the data obtained 
on the short spatial scales. The CARMA and SMA observations achieve a 
noise level of 0.5 mJy/beam and 2.6 mJy/beam, respectively.

Flux densities integrated on a circular aperture of 1.5\arcsec\ in diameter are
53$\pm$2 mJy and 204$\pm$8 mJy at 1.3~mm and 0.88~mm respectively, 
where the uncertainties reflect random noise only. The spectral index measured 
between 0.88~mm and 1.3~mm is 3.6$\pm$0.5(random noise)$\pm$1.0
(flux calibration), which is consistent within the large uncertainty with the
mean value of 2.6 measured toward nearby circumstellar disks \citep[][]{Ricci10a}.

The CARMA and SMA maps show similar features: 
the dust emission from the disk extends for about 1\arcsec\ in diameter 
(250 AU), and has a local minimum at the stellar position, which is 
indicated by the star symbol in Figure 1 (the size of the symbol is 
the same of the uncertainty on the position of the star).
Both maps reveal that the west side of the disk is brighter than 
the east side, though this asymmetry is more evident at 1.3 mm than at 0.88~mm. 
The 1.3~mm intensity has a maximum of 9.3$\pm$0.5 mJy/beam 
located 0.4\arcsec\ westward from the star, while the value at the symmetric 
point is only 2.8$\pm$0.5 mJy/beam. 
The integrated flux density westward and eastward of the star is 
35$\pm$1 mJy and 18$\pm$1 mJy, respectively.
At 0.88 mm, the integrated flux densities for the west and east side of the disk 
are 109$\pm$4 mJy and 94$\pm$4 mJy respectively, with a peak of 33.0$\pm$2.6 mJy/beam in the 
west side and 26.1$\pm$2.6 mJy/beam in the symmetric position.

\subsection{Fourier domain}
\label{sec:asym}

A better understanding of the morphology of the dust emission observed 
toward LkH$\alpha$~330 can be achieved by inspecting the complex visibilities, 
which, contrary to the intensity maps, are not affected by the synthesized 
beam smoothing and by possible artifacts introduced during the image 
deconvolution process. 

\begin{figure}[!t]
\centering
\includegraphics[angle=0, width=1.2\linewidth]{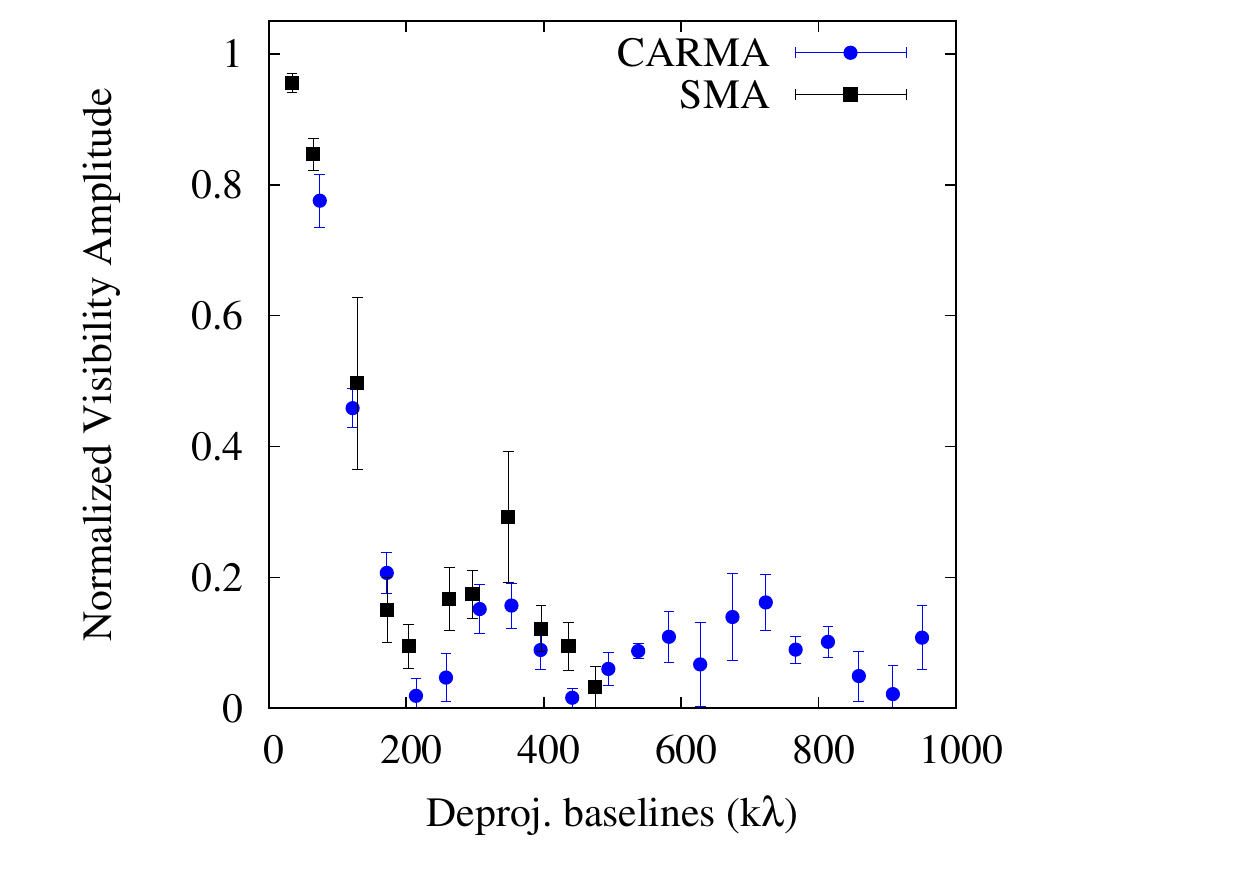}
\caption{\label{fig:vis} Normalized visibility amplitude vs. deprojected 
baseline. CARMA  and SMA data are shown with open circles and filled squares respectively. }
\end{figure}

Figure~\ref{fig:vis} shows the circularly averaged CARMA and SMA 
visibility amplitude vs. the baseline length, where 
this latter quantity has been deprojected assuming a disk inclination 
of 35\arcdeg\ and a position angle of 80\arcdeg\ as derived by \cite{Andrews11}.
To allow a comparison between CARMA and SMA data, the visibility amplitude has 
been normalized by the integrated flux density at the respective wavelengths.

The visibility profile drops to zero at about 200 k$\lambda$, implying 
that the emission is resolved on angular scales of about 1\arcsec. 
The presence of two additional nulls, 
located at  450~k$\lambda$ and 900~k$\lambda$ respectively, indicate 
that the surface brightness is characterized by sharp radial variations.
\cite{Andrews11} modeled the SMA observations 
with a circular symmetric inclined disk characterized by a partially 
dust-depleted cavity of 70 AU in radius.  In this model, the dust surface 
density has a discontinuity at the outer cavity radius, which produces 
the first two lobes of the visibility profile. For the rest 
of the paper we will assume this as a reference disk model and will  focus 
the discussion on the asymmetry observed in the dust emission. 

While the visibility amplitude probes the angular scales of the observed 
emission, the imaginary part of the visibility (called {\it imaginary visibility} 
henceforth) provides information 
about its symmetry properties with respect to the phase center of the observations, 
i.e., the position in the sky at which the telescope was pointed. 
The Fourier transform has the property that a purely real even 
function, such as a two-dimensional centro-symmetric disk surface brightness, 
maps in a purely real function. Therefore, nonzero imaginary visibilities, 
which corresponds to the imaginary part of Fourier transform of the disk surface 
brightness, indicate the presence of an asymmetry with respect to the phase center 
of the observations, which, in the case of LkH$\alpha$~330 observations, corresponds
to the expected position of the star corrected for the stellar proper motion (see Section~2). 

\begin{figure*}[!t]
\includegraphics[angle=0, width=0.33\linewidth, bb= 30 0 320 290, clip=true]{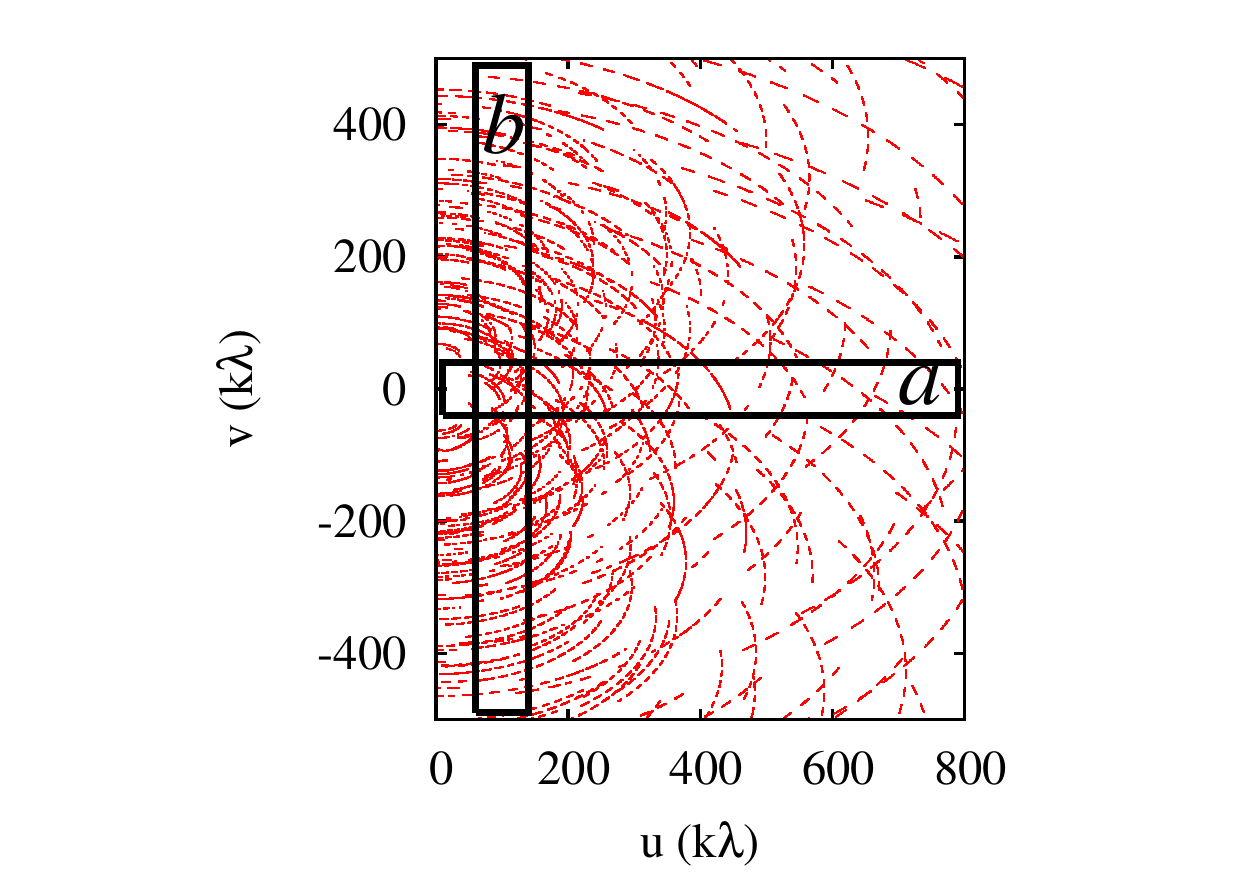}
\includegraphics[angle=0, width=0.33\linewidth, bb= 30 0 320 290, clip=true]{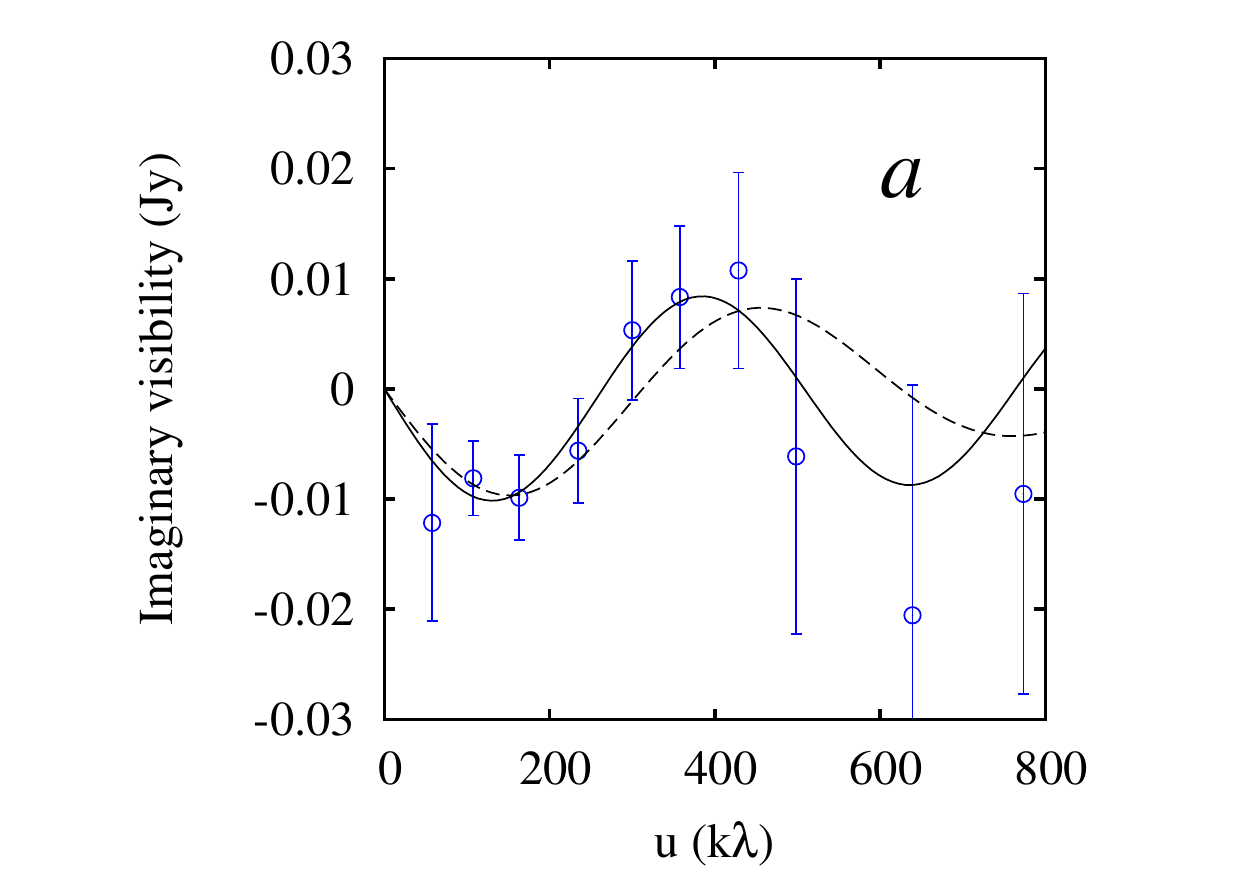}
\includegraphics[angle=0, width=0.33\linewidth, bb= 30 0 320 290, clip=true]{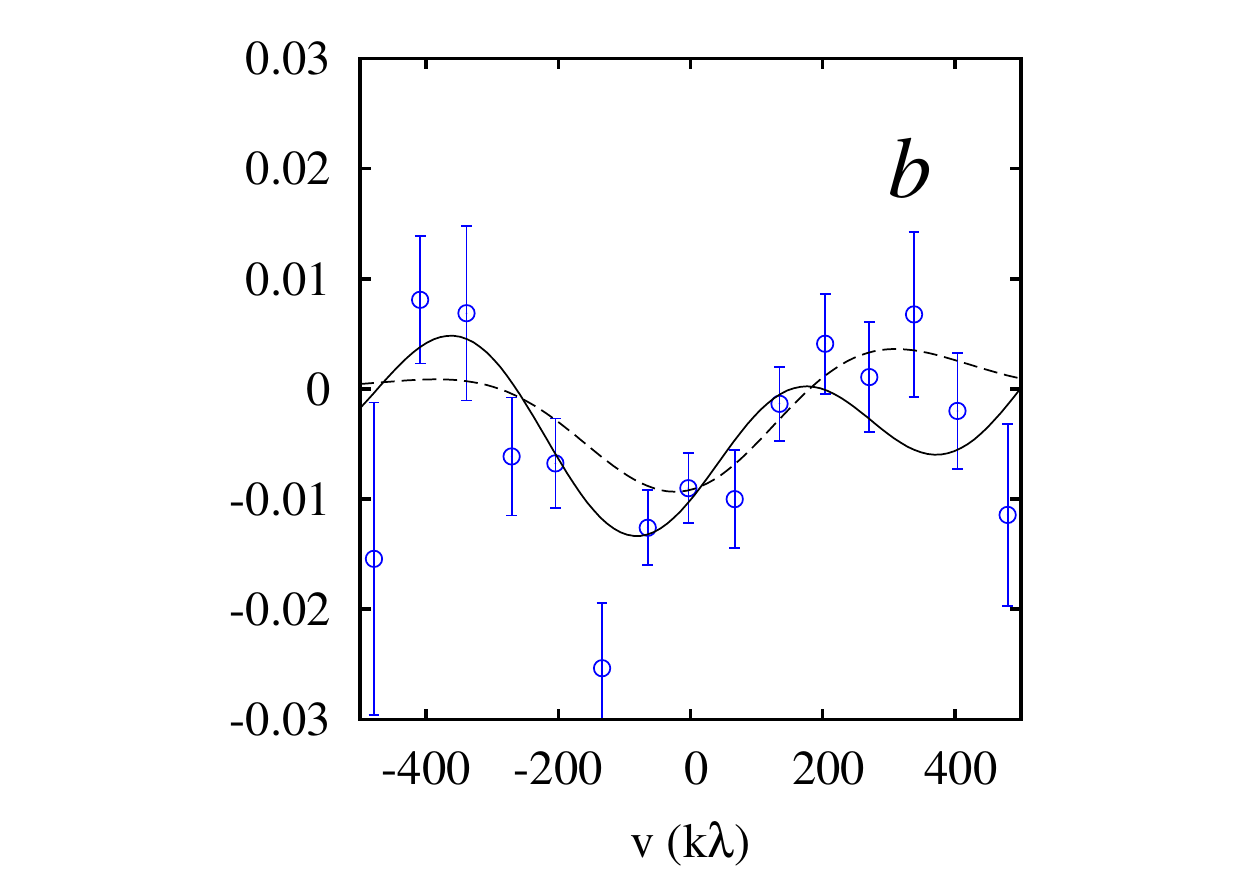}
\caption{\label{fig:imag_carma} Left: coverage of the  {\it uv} plane of the CARMA 
observations of LkH$\alpha$~330. Center:  imaginary visibilities 
measured on the {\it uv} plane region labeled as $(a)$. Open circles indicate 
values averaged between $-40$~k$\lambda$ and 40~k$\lambda$ along $v$
and every 80~k$\lambda$ along $u$. Right: imaginary part of the correlated flux measured 
on the {\it uv} plane region labeled as $(b)$, averaged from 60~k$\lambda$ and 140~k$\lambda$ 
along $u$ and every 80~k$\lambda$ along $v$. 
The dashed and solid black curves show the imaginary visibilities of the single- and two- components 
best fit models, whose properties are listed in Table 1.}
\end{figure*}

Figure~\ref{fig:imag_carma} presents the {\it uv} coverage achieved by CARMA 
observations and the imaginary visibilities along $\it u$ 
and $\it v$, as calculated by binning the data every 80~k$\lambda$.  Although the 
following analysis makes use of all the {\it uv} data, for sake of clarity we only show the 
values measured inside the rectangles $\it (a)$ and $\it (b)$.
In these regions, the imaginary visibilities clearly deviate from zero, and oscillate 
around zero reaching maximum values around 15 mJy. This  suggests that about 30\% of
the integrated flux arises from asymmetric structures. 

In the following, we analyze the imaginary visibilities 
to constrain the geometrical properties  of the asymmetries in the disk emission. To this end, 
we assume that any arbitrary asymmetric structure in the image domain can be expressed as a 
combination of  two-dimensional (2-D) Gaussian functions, 
\begin{multline}
g(x,y)= A \times  \exp \left\{ - \frac{ [(x-x_0)\cos\theta+(y-y_0)\sin\theta]^2   }{2\sigma_{maj}^2} \right\} \times\\
	\times \exp \left\{ -\frac{[(x-x_0)\sin\theta-(y-y_0)\cos\theta]^2}{2\sigma_{min}^2} \right\}, 
\end{multline}
where $A$ is the intensity at the central position ($x_0$,$y_0$), 
$\sigma_{maj}$ and $\sigma_{min}$ are the dispersions along the 
major and minor axis, and $\theta$ is the position angle  of the major axis with respect 
of the $x$ axis measured counter-clockwise. 
The Fourier transform of Equation 1 can be calculated analytically (see the Appendix), and 
its imaginary part can be expressed as
\begin{multline}
\Im(G(u,v))= A \times 2\pi \sigma_{maj} \sigma_{min} \times \sin[-2\pi(u x_0 + v y_0)] \times  \\
	\times  \exp\{-2\pi^2 [ (u\cos\theta+v\sin\theta)^2\sigma_{maj}^2+\\ 
		+(-u\sin\theta+v\cos\theta)^2\sigma_{min}^2 ] \}.
\end{multline}
Note that $\Im(G(0,0))= A \times 2\pi \sigma_{maj} \sigma_{min}$ is the total flux of 
a 2-D Gaussian functions.

To analyze the observations, we start by assuming a single Gaussian component 
and calculate the best fit values for  $x_0$, $y_0$, $\sigma_x$, $\sigma_x$, $A$, $\theta$, 
as well as the corresponding uncertainties, through a nonlinear least-squares 
Marquardt-Levenberg fit of the imaginary visibilities measured by CARMA. 
If the reduced $\chi^2$, $\chi^2_r$, obtained with a single component is significantly 
larger than unity, we repeat the fitting procedure assuming two Gaussian components. 
We continue adding Gaussian components until $\Delta\chi^2_r = \chi^2_r-1$ is less than the value 
corresponding to the 99\% confidence level agreement between the model and the observations. 
This latter value depends on the number of degrees of freedom in the model fitting, 
as well as on the number of visibilities, and, in the case of CARMA observations, 
is 0.04, 0.07, and 0.1 for one, two, and three Gaussian components, respectively\footnote{These values are
calculated as $\Delta\chi^2/(N-p)$ where $N=282$ is the number of visibilities, 
$p=$6, 12, and 18  is the number of free parameters corresponding to 1, 2, and 3 Gaussian 
components respectively, and $\Delta\chi^2$=10.64, 18.55, and 25.99 is the value  
corresponding to 99\% confidence level in the case of 6, 12, and 18 free parameters.}.  

\begin{figure*}[!t]
\includegraphics[angle=0, width=0.9\linewidth]{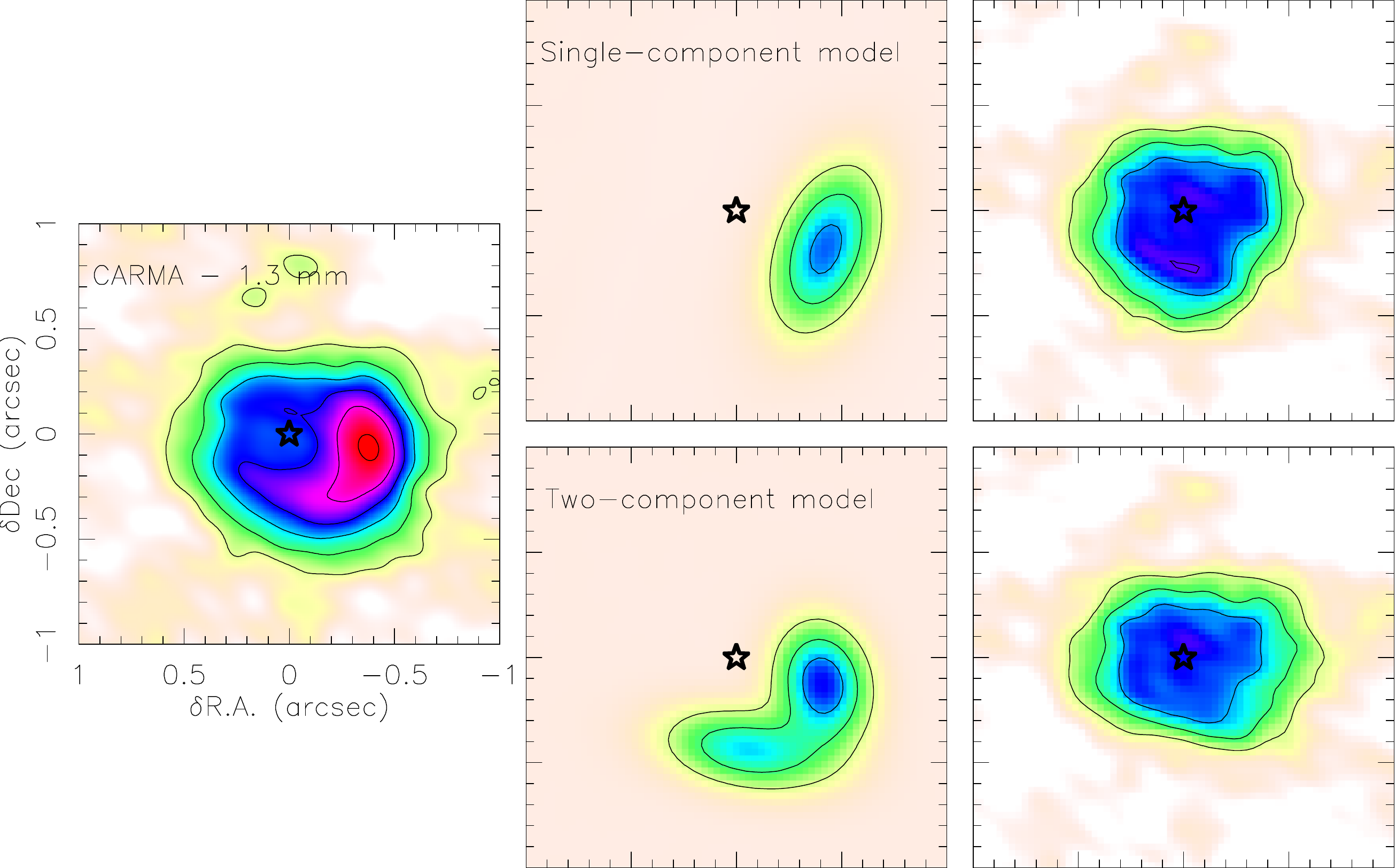}
\caption{\label{fig:model} Left: map of the dust continuum emission observed by CARMA at a wavelength of 1.3 mm 
toward LkH$\alpha$~330. The color scale and the contours are as in Figure 1. Center: maps of 
the best fit models for the imaginary visibilities. Right: maps of the residuals obtained by 
subtracting in the Fourier space the best fit model for the imaginary visibilities from 
the observations. The single- and two-component best fit models are shown in the top and 
bottom panels, respectively.}
\end{figure*}

We find two Gaussian components are required to obtain a good fit 
of the imaginary visibilities ($\chi_r^2=1.03$), while the best fit model with 
a single component is outside the 99\% confidence level ($\chi_r^2 = 1.09$).
The properties of both the single- and two-component best fit 
models are summarized in Table 1, while Figure~\ref{fig:imag_carma} shows 
the comparison in the Fourier space between the best fit models and the 
observations.
As a check of the fitting procedure, we decomposed the observed 
1.3~mm dust emission into its asymmetric and symmetric components.
Figure~\ref{fig:model} shows the 1.3 mm CARMA observations, 
the maps of single- and two-component best fit models,
and the residual maps obtained by subtracting in the Fourier space the models 
from the observations. The residual map for the single-component model 
shows a faint residual asymmetry in the south part of the disk, which is accounted for 
by the two-component best fit model. The appearance of asymmetries in the residual map 
is consistent with the $\chi^2$ analysis, giving us confidence on the procedure employed to 
identify asymmetric structures in the LkH$\alpha$~330 disk.

The two Gaussian components that fit the imaginary visibilities have integrated fluxes 
of 9 mJy and 7 mJy, respectively, which account for about 30\% of the total flux density 
measured at 1.3 mm. They are centered at a radius of about 0.42\arcsec\ from the central star, 
which corresponds to a separation of about 100 AU, and are oriented so 
that the minor axis is aligned along the radial direction and the major axis is tangential 
to a circle of 100 AU in radius. Along the major axis, they extend by more than 70 AU, while  
along the minor axis they appear to be spatially unresolved. This sets the upper limit for 
their radial extent to about 25 AU.

\begin{table*}[!t]
\caption{\label{tab:model} Gaussian model}
\centering
\begin{tabular}{lccccccc}
\hline
\hline
& \footnotemark$x_0$ (\arcsec) & $y_0$ (\arcsec)& $\sigma_{maj}$ (\arcsec) & $\sigma_{min}$ (\arcsec) & $F$ (Jy) & $\theta$ (\arcdeg) & $\chi^2_r$ \\ 
\hline
\multicolumn{7}{l}{Single-component best fit model} & 1.09 \\
g1 & 0.37$\pm$0.03 & -0.20$\pm$0.03 & 0.16$\pm$0.06 & $<0.05$ & 0.011$\pm$0.002 & 70$\pm$5 &   \\
\hline
\multicolumn{7}{l}{Two-component best fit model} & 1.03 \\
g1 & 0.40$\pm$0.03 & -0.11$\pm$0.02 & 0.14$\pm$0.05 &  $<0.05$ & 0.009$\pm$0.002 & 88$\pm$5 &     \\
g2 & 0.05$\pm$0.03 & -0.43$\pm$0.03 & 0.21$\pm$0.07 & $<0.05$ & 0.007$\pm$0.002 & 8$\pm$5  &    \\
\hline
\end{tabular}
\footnotetext{The $x$ coordinate is positive toward west, i.e., $dx = -d\textrm{RA}$}
\end{table*}

\begin{figure*}[!t]
\includegraphics[angle=0, width=0.33\linewidth, bb= 30 0 320 290, clip=true]{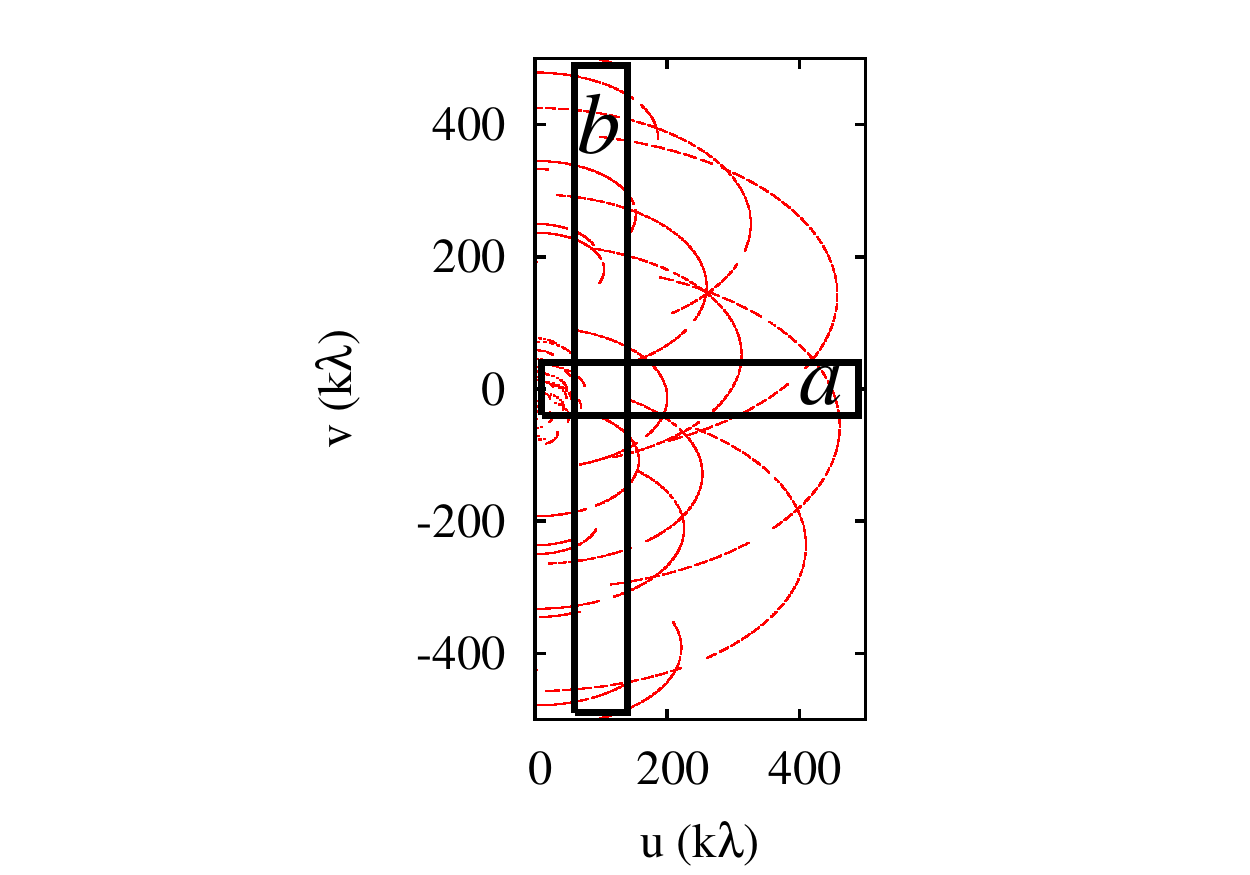}
\includegraphics[angle=0, width=0.33\linewidth, bb= 30 0 320 290, clip=true]{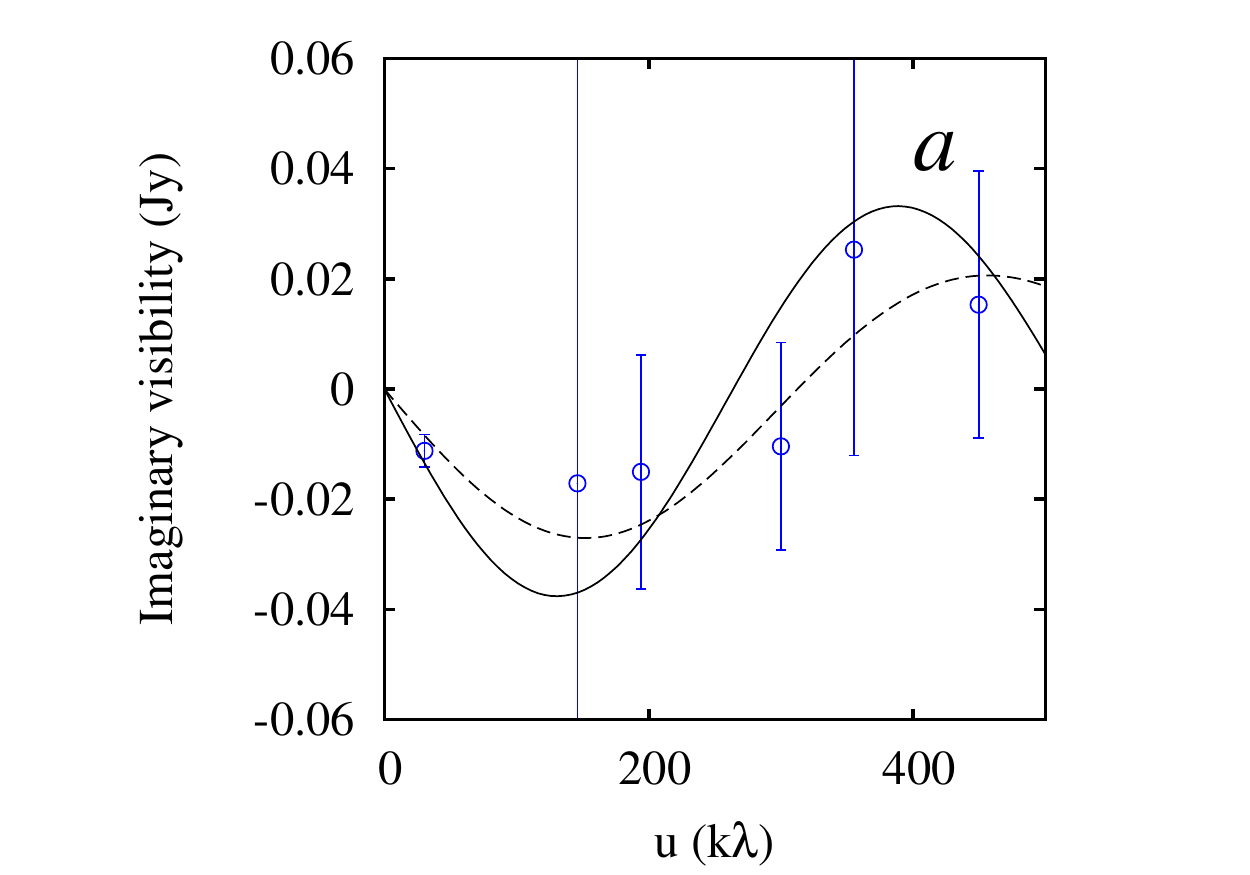}
\includegraphics[angle=0, width=0.33\linewidth, bb= 30 0 320 290, clip=true]{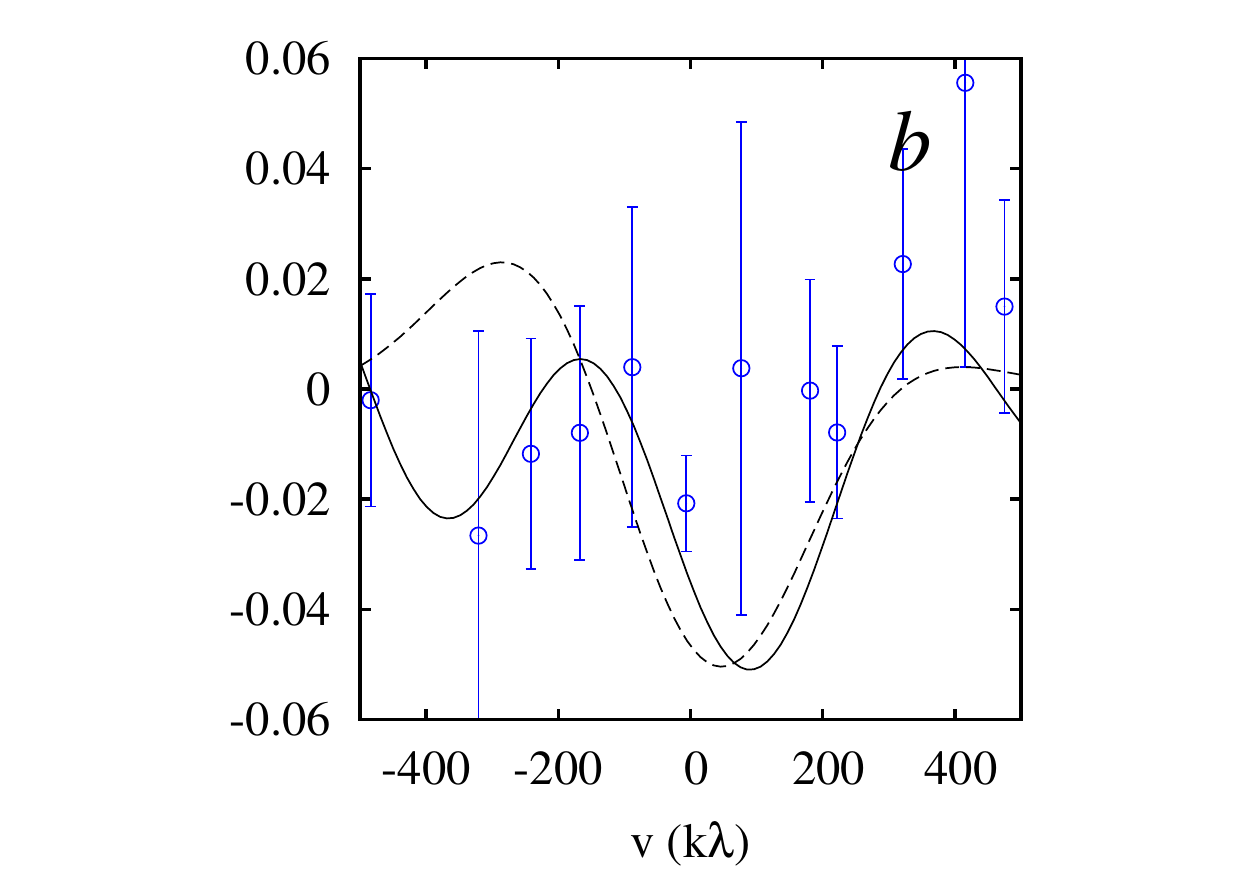}
\caption{\label{fig:imag_sma} Left: coverage of the {\it uv} plane of the SMA observations of LkH$\alpha$~330. 
Center and right: imaginary visibilities measured as in Figure~\ref{fig:imag_carma}.
The dashed and solid black curves show the imaginary visibilities of the single- and two- components 
best fit models, where the flux of the asymmetric components has been scaled 
to account for the difference in  wavelength by assuming a spectral index $\alpha=3.6$, 
as measured from the integrated fluxes at 0.88~mm and 1.3~mm.}
\end{figure*}

Figure~\ref{fig:imag_sma} shows the {\it uv} coverage achieved by the SMA 
observations, as well as the imaginary visibilities calculated 
as in the case of the CARMA data. We find that, due to the sparse {\it uv} coverage 
for spatial frequencies larger than 100~$k\lambda$,  the SMA data provide poor 
constraints on disk asymmetries. More specifically, the SMA imaginary visibilities 
are consistent within 99\% confidence level with both a symmetric model for the dust
emission and with the single- and two-component best fit Gaussian models derived 
from the CARMA observations. These latter are shown with dashed and solid black 
lines, respectively, and have been calculated by assuming an optically thin emission 
with a spectral index $\alpha = 3.6$. Lower values of $\alpha$ will lead to a lower 
flux of the asymmetric structures at 0.88~mm and therefore to an even fainter signal in 
the imaginary visibilities. By contrast, the imaginary visibilities of the asymmetries inferred
at 1.3 mm would be incompatible with the measured values at 0.88~mm for $\alpha>4$. 
This is however an unlucky case since it would imply a dust opacity slope $\beta$ larger 
than 2.

\section{Discussion}
\label{sec:disc}

Models of the SMA observations of LkH$\alpha$~330  find that the 
0.88~mm dust thermal emission is optically thin, with a maximum optical depth of 
about 0.3 at the outer edge of the dust-depleted cavity located at 70 AU 
from the central star \citep{Andrews11}. Since dust opacity 
of typical interstellar grains decreases with the wavelength \citep[i.e., $\beta \geq 0$,][]{Draine06}, 
then the 1.3~mm dust emission from LkH$\alpha$ disk should also by optically thin. 
We can therefore express it as
\begin{equation}
\label{eq:mm_flux}
I_\nu(r,\theta) \propto B_\nu(T(r,\theta)) \times k_\nu(r,\theta) \times \Sigma_g(r,\theta),
\end{equation}
where $T$ is the temperature of the disk mid plane, $k_\nu$ is the dust opacity {\it per gram of gas}, 
and $\Sigma_g$ is the gas disk surface density. The asymmetries observed in 
the dust emission can therefore be caused by the deviation from the central symmetry 
of the disk density, temperature, and/or opacity. 
In the following we discuss possible physical effects that might lead to such variations.

\subsection{Azimuthal density perturbations due to the disk-planet interaction}
\label{sec:disc_inter}

Planets embedded in nearby circumstellar disks are 
too faint to be directly observed at millimeter wavelengths. 
However, planets 
more massive than Jupiter will gravitationally interact with the surrounding 
circumstellar material, leading to the formation of annular gaps and density waves 
in the gas density distribution around the planetary orbit
\citep{Lin86,Bryden99,Rafikov02,Crida06}. These structures might 
have spatial scales of tens of AU and can be observed by existing telescopes, therefore 
 providing an indirect tool to study the formation of planets. 

Dynamical clearing by giant planets has been proposed as a possible mechanism 
to explain the lack of near-infrared emission and the presence of millimeter cavities 
toward transitional disks \citep{Brown08,Isella10b,Andrews11,Isella12}. 
\citet[hereafter DS11]{Dodson11} suggest that planetary systems 
composed of two or more giant planets might naturally explain millimeter cavities
larger than 15 AU. Similar results are found by \cite{Zhu11}, who also argue 
that multiple giant planets would lead to a mass accretion rate onto the central star 
which is orders of magnitude lower than that measured toward transitional disks with 
large millimeter cavities.  
In a following paper, \cite{Zhu12} suggest that one single 
giant planet orbiting close to the outer edge of the millimeter cavity might be sufficient 
to explain the cavity diameter if millimeter size grains  are trapped in the pressure 
bump that forms at the outer edge of the dust-depleted cavity \cite[see also][]{Pinilla12}.

While it is generally accepted that one or more giant planets can lead to large millimeter 
cavities, the role of the disk-planet interaction in creating azimuthal asymmetries in the 
disk emission has not been explored in great detail. In the case of a single Jupiter mass planet, 
\cite{valborro07} find that large scale vortices can form at the outer edge of the annular 
gap cleared by the planet as a result of Rossby wave instabilities. 
Azimuthal asymmetries in the disk surface density also appear in hydrodynamic simulations 
with multiple giant planets by DS11, but it remains to be shown if they are a general 
outcome of the disk-planet interaction, how they depend on the orbital parameters and 
mass of the planets, and  how they will influence the millimeter-wave dust thermal emission.

To address these questions, we have performed hydrodynamic 
simulations using FARGO,  a publicly available, polar grid-based 2-D hydrodynamic code expressly 
designed to study the disk-planet interaction \citep{Masset00}.
To date, FARGO has been employed to investigate the radial migration of planets in viscous 
disks (Masset 2000, 2001, 2002; Masset \& Papaloizou 2003; Masset \& Ogilvie 2004; Masset et al. 2006;
Masset \& Casoli 2009, 2010; Crida et al. 2007) and the effects of circumplanetary disks on planet
migration (Crida et al. 2009). FARGO was also employed to study the opening of annular gaps in
the disk surface density due to the gravitational field of a planet (Crida et al. 2006) and to check
whether forming planetary systems might be able to open cavities as large as of those observed in
transitional disks (DS11; Zhu et al. 2011).

The simulations are performed adopting a procedure similar to that described in DS11. 
 First, we generate a planetary system that is able to  
carve a cavity characterized by an outer radius of 70 AU, as observed in the LkH$\alpha$~330 disk \citep{Andrews11}. 
To do that, we assume that the size of the gap opened by the gravitational 
torque exerted by a planet on the disk is five times the 
planet Hill radius, defined as $R_H = R_p (M_p/3M_\star)^{1/3}$, where $R_p$ and 
$M_p$ are the planet's orbital radius and mass respectively \citep[see, e.g.,][]{Bryden99}.
To create a cavity of 70 AU in radius, the outermost planet should therefore have an 
orbital radius equal to $70 \textrm{AU} \times (1+2.5(M_p/3M_\star)^{1/3})^{-1}$. 
Gaps created by multiple planets with the same mass will overlap if 
their orbital radii are separated by $\Delta r_{ij} = 2.5 (r_i+r_j)  (M_p/3M_\star)^{1/3}$.
In our reference model, we assume a planet mass of  5 M$_J$ and, following the previous relations, 
we place four planets at orbital distances of 55 AU,  34~AU, 21 AU, and 13 AU to have a cavity
 from 10 AU to 70 AU. We anticipate here that the mass and number of planets do not 
 affect the main results of our analysis.

 The disk properties are as in DS11: we assume a constant aspect 
ratio $h_p/r=0.05$, where $h_p$ is the disk pressure scale height; 
the disk surface density is expressed by a power law $\Sigma \propto r^{-1}$; 
the disk viscosity is parameterized using the Shakura-Sunyaev $\alpha$ parameter 
and we assume $\alpha = 0.002$. 
The inner and outer radius of our simulations are set to 1 AU and 160 AU respectively, 
and we choose the surface density normalization 
constant so that the disk mass is 0.03~M$_\sun$, i.e., similar to that inferred from 
SMA observations. Finally, as in DS11, we allow the planets to feel the torque 
from the disk after  they have opened a gap in the disk surface density. This 
artificially suppresses the Type I migration which is not properly described 
by the FARGO code and would cause rapid inward migration of the planets.

\begin{figure*}[!t]
\centering
\includegraphics[angle=0, width=\textwidth]{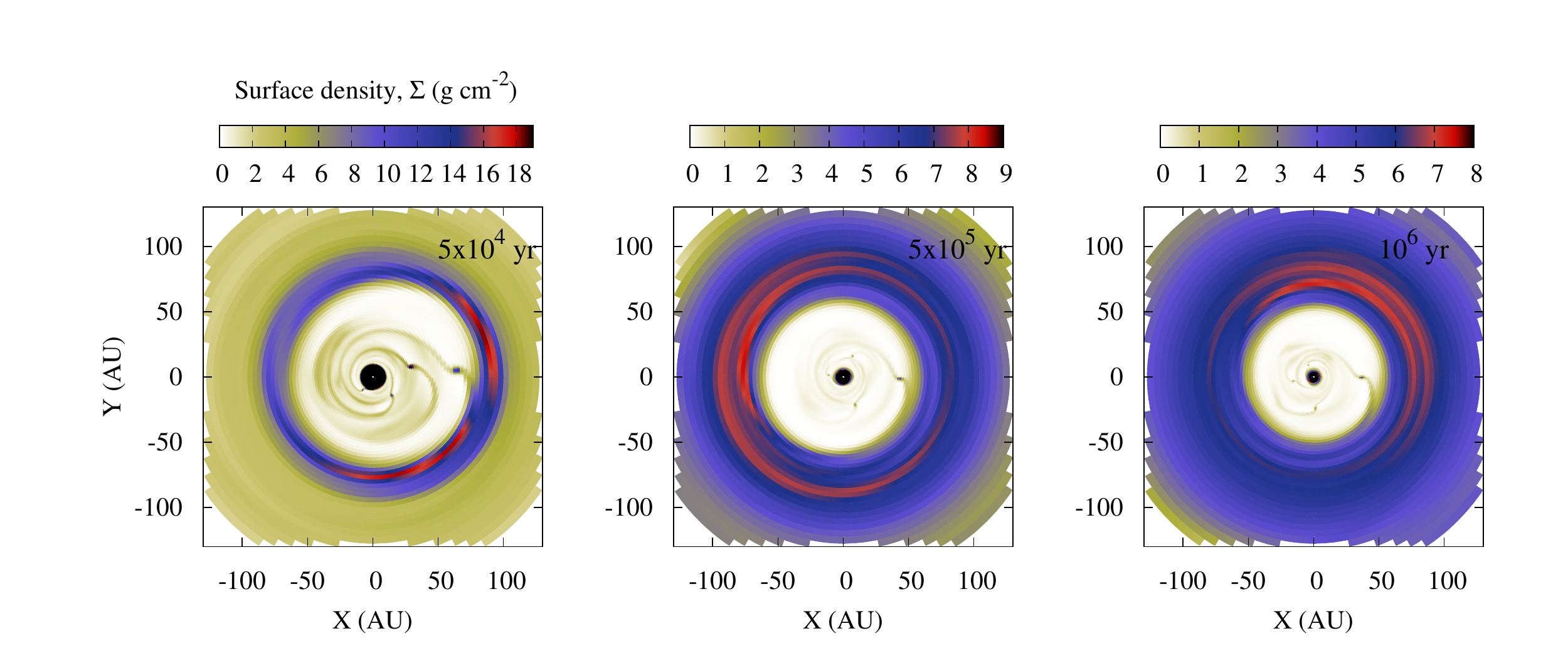}
\caption{\label{fig:fargo} Hydrodynamic simulations of the perturbing effects of a planetary 
systems composed by four 5~M$_J$ planets on the disk surface density. The three panels
from the left to the right show snapshots of the surface density after $5\times10^4$ yr, $5\times10^5$ yr, and 
$10^6$ yr respectively.}
\end{figure*}

Figure~\ref{fig:fargo} shows the gas surface density map after 100, 1000, and 2000 orbits of 
the outermost planets, which corresponds to about $5\times10^4$ yr,  $5\times10^5$ yr, and $10^6$ yr 
respectively. A cavity of about 70 AU in radius is created after a few orbits of the outermost 
planet. Between about 70 AU and 90 AU, azimuthal asymmetries with the shape 
of circular arcs develop. After $5\times10^4$ yr, the surface density at 80 AU from the star 
varies azimuthally by more than a factor of two, from about 18 g cm$^{-2}$ to about 8 g cm$^{-2}$. 
This density contrast has about the same amplitude of the intensity contrast measured in 
the 1.3~mm dust emission, as discussed in Section~\ref{sec:obs}. In addition, the bright circular 
arcs have roughly gaussian density profiles both radially and azimuthally with dispersions 
of 10 AU and 50 AU respectively, which are similar to size of the asymmetries derived from the
CARMA observations (see Section~\ref{sec:asym}).  Snapshots of the surface density at later ages 
show that the azimuthal asymmetries between 70 AU and 90 AU are persistent, although the 
density contrast decreases with time down to about 30\% after $10^6$ yr.
In the assumption that dust and gas are well coupled (we investigate this assumption in the next section),
these simulations indicate that disk-planet interaction might naturally explain the morphology of the dust 
emission observed toward LkH$\alpha$~330, 
and, if taken at face value,  would suggest that an azimuthal contrast of about 2 in the millimeter dust emission 
might result from a recent ($<10^4$~yr) episode of planet formation. 

However, the geometrical properties and the temporal evolution of the azimuthal 
asymmetries observed in hydrodynamic simulations might depend on a number of unknown, or 
poorly constrained parameters, such as  the mass, the orbital radius, and number of 
planets, and the disk viscosity. An extensive exploration of the parameter space 
is therefore required to constrain the properties of possible companions orbiting LkH$\alpha$~330
from the existing millimeter-wave observations.  Such a study requires a large 
amount of computational time, which goes beyond our present computational capabilities 
and the scope of this work. In this paper, we limit our analysis to investigate whether the formation 
of azimuthal asymmetries in the disk surface density is a general outcome of the disk-planet 
interaction or, on the contrary, requires a fine tuning of the model parameters. 
To this end, we performed short ($10^4$ yr) hydrodynamic simulations using 
FARGO, in which we explored cases with one, two, and three planets characterized 
by masses between 3 M$_J$ and 10 M$_J$, and deployed so that the outer edge of the 
cleared cavities is located at about 70 AU as derived for LkH$\alpha$~330. We 
varied the disk viscosity from $\alpha$=0.002 and $\alpha$=0.02,  and changed the 
disk aspect ratio $h_p/r$ to account for a possible flaring geometry.
In total we performed about 20 simulations. In all these cases, we  
observed that the disk-planet interaction leads to large scale azimuthal asymmetries 
in the disk surface density between 70~AU and 90~AU from the central star. 
At the end of our simulations, i.e., after about $10^4$ yr, the density contrast at 80 AU 
varies azimuthally between a few \% to almost a factor of 10. The largest variations are 
observed in simulations with planets of 10 M$_J$ and low disk viscosity. 
Furthermore,  we find that the azimuthal asymmetries are fainter and disappear in a 
shorter time scale in the case of high disk viscosity. These preliminary results (a more
extensive exploration of the parameter space will be presented in a future paper)
suggest that wide separation companions more massive than a few Jupiter 
masses might quite naturally explain the morphology of the dust emission 
observed toward LkH$\alpha$~330, while less massive planets 
might not produce sufficiently strong asymmetries.

A recent direct-imaging survey has found that less than 
20\% of debris disk stars have a planet more massive than 3~M$_J$ beyond 10 AU \citep{Wahhaj13}, 
suggesting that planetary systems like those proposed above are rare. 
However, circumstellar disks characterized by cavities larger than 15 AU are also rare, 
with a frequency less that 10\% among the entire disk population \citep{Andrews11}.
The paucity of wide separation massive planets is therefore not in contrast with
the suggested explanation for the morphology of LkH$\alpha$~330 disk.

\subsection{Azimuthal density perturbations due to disk instabilities}
\label{sec:disc_rosby}

Rossby-wave and baroclinic instabilities have been proposed as a possible 
mechanism to form lopsided asymmetries in the millimeter dust continuum emission 
without the need of companions \citep{Klahr03,Wolf02}.

\cite{Regaly12} suggested  that a large-scale vortex triggered by 
Rossby wave instability might be responsible for the asymmetry observed in the
LkH$\alpha$~330 dust continuum emission at 0.88 mm \citep{Brown08}.
In their model, the instability is triggered by a discontinuity in the disk viscosity 
due to the transition from a magneto-rotational instability (MRI) dead-zone to an MRI 
active zone. We note that this hypothesis faces two major problems. First, 
the dead-zone should extend up to a distance of 70 AU from the central star, which 
is much larger than the extent predicted by theoretical models \citep[see, e.g., the review by][]{Armitage11}. 
Second, Rossby wave 
instabilities  do not necessarily lead to a depletion of gas and dust inside the dead-zone region.
The gas depletion preferentially happens in the active disk region, i.e. at 
distances larger than 70 AU,  as the gas migrates inward and it is trapped into the 
vortex (see Figure 4 of Reg\'aly's paper). This model then may not explain the lack of 
infrared excess and the spatially resolved observations at 
millimeter wavelengths. A possible work around is that mm-size particles might 
be trapped in the vortex located at the outer edge of the dead-zone, resulting in a 
dust filtration process similar to that discussed by Zhu et al. (2012). This would 
lead to a reduced millimeter-wave dust opacity inside the dead-zone. However, 
dust particles smaller that about 1~$\mu$m might remain coupled to the gas and 
produce a significant amount of near-infrared emission. 

Circumstellar disks heated by the central star are characterized by a radial gradient in the disk 
temperature which makes them prone to baroclinic 
instabilities \citep{Raettig13}. Earlier three dimensional numerical simulations, 
which were characterized by a very small radial dynamic range, have shown that baroclinic 
instabilities lead to the formation of anticyclonic vortices, which might eventually merge 
and produce azimuthal structure similar to that observed toward LkH$\alpha$~$330$ 
\citep{Klahr03}. In the last ten years, several theoretical investigations have expanded 
on this result to study the formation
and evolution of vortices in rotating circumstellar disks, sometimes leading to controversial 
results \citep[see, e.g.,][]{Johnson06}. Although a three-dimensional model 
that covers the radial extent of a disk, i.e., a few hundred AU, is still missing, the most 
recent simulations confirm that vortices induced by baroclinic instabilities can form and survive for 
several hundred orbits \citep[e.g.,][]{Petersen07a,Petersen07b,Lesur10, Raettig13}.
However, it has not yet been explored whether this process can produce vortices 
in the outer regions of circumstellar disks, where baroclinic instabilities might not exist due to the fact that 
disks are almost isothermal in the radial direction.

\subsection{Azimuthal variations in the dust opacity}

Equation~\ref{eq:mm_flux} shows that azimuthal variations in the millimeter-wave 
dust emission can also result from azimuthal variations in the dust opacity.  
This might be the case if, for example, large 
dust grains are trapped in maxima of the gas pressure caused by 
local enhancements in the gas density.
This process has been suggested to be involved in the formation of planetesimals, 
and it has been proposed as a possible solution for the radial drift problem of large 
particles in a turbulent disk \citep{Pinilla12}. 
More recently, \cite{Birnstiel13} have studied the dust trapping caused by 
azimuthal asymmetries in the gas surface density that might result, for example, 
from disk-planet interactions. They find that, if the density asymmetries are 
long-lived in the gas corotating rest frame, even weak variations of the order of 10\% 
in the azimuthal gas density can lead to a large concentration of mm-size dust grains 
toward the peak of the gas density. Since the dust opacity at mm wavelengths is 
dominated by mm-size particles, dust trapping might therefore be a dominant process 
in the formation of asymmetries in the millimeter-wave dust continuum emission
\citep{Marel13}.

The efficiency of dust trapping depends critically on the stability of the pressure 
bumps and on the time scale required to concentrate particles. 
The stability of pressure bumps created by the disk-planet interaction can be investigated by analyzing 
the temporal evolution of the azimuthal gas asymmetries as calculated using FARGO. 
For the reference model shown in Figure~\ref{fig:fargo}, three phases can be identified. 
The initial phase lasts for about $10^{4}$ yr, and is very dynamic. Azimuthal asymmetries form 
and merge on a few orbital time scales, and no clear rotational pattern can be identified.
A second phase starts after about $10^{4}$ yr, when a main lopsided asymmetry appears 
outside the gas-depleted cavity at an orbital radius of about 80 AU. The rotational velocity of this structure 
is initially about 70\% of the local Keplerian velocity and increases with time up to 100\% of 
the Keplerian velocity after about  $10^5$ yr.
In the third phase, which lasts from $10^5$ yr to the end of the simulation at about $10^6$ yr, 
the lopsided asymmetry is co-rotating with the local gas at Keplerian speed. 
These results suggest that planet-induced asymmetries might be stable for a long time.

Following \cite{Birnstiel13}, dust particles can be concentrated around gas pressure maxima 
only if $St \geq \alpha$, where $St \simeq \rho_s a/\Sigma_g$ is the Stoke number 
of a spherical dust grain of radius $a$ and internal density  $\rho_s$, $\Sigma_g$ is the local
gas surface density, and $\alpha$ is the Shakura-Sunyaev viscosity parameter. 
For $\alpha=10^{-3} $--$10^{-2}$ 
\citep[][and references therein]{Armitage11} and $\rho_s=2$ g cm$^{-3}$ \citep{Pollack94}, the 
concentration of millimeter-size dust grains therefore requires surface densities less than 
$20-200$ g cm$^{-2}$. This condition is generally satisfied in the outer regions of circumstellar 
disks \citep[see, e.g.,][]{Isella09}.
The time scale required to concentrate the dust grains can then be expressed as
\begin{equation}
\tau_c \sim \frac{1}{\rho_s a} \left( \frac{1}{\Sigma_g^{min}}-\frac{1}{\Sigma_g^{max}} \right)^{-1}  
\left( \frac{	h_p}{r} \right)^{-2} \frac{2\pi}{\Omega_k}
\end{equation}
where $\Omega_k$ is the angular Keplerian velocity,  and $\Sigma_g^{min}$ and 
$\Sigma_g^{max}$ are the minimum and maximum values of the gas disk surface
density at the orbital radius $r$.
In the specific case of the LkH$\alpha$~330, we calculate that the time scale to concentrate 
dust grains of 1 mm and 1~cm at a distance of 80 AU from the central star would be 
of about 10 Myr and 1 Myr, respectively, for $\Sigma_g^{min}=8$ g cm$^{-2}$, 
$\Sigma_g^{max} = 18$ g cm$^{-2}$, and $h_p/r=0.05$ as in the hydrodynamic simulations 
shown in the left panel of Figure~\ref{fig:fargo}. Although these estimates might 
be uncertain by a factor of several due to the uncertainty on the gas surface density,
they suggest that the concentration of large grains in LkH$\alpha$~330 outer disk 
might require a temporal scale comparable with the age of the stellar system.

As proposed by \cite{Birnstiel13}, spatially resolved observations of the dust thermal 
emission at mm and cm wavelengths will measure azimuthal variations in the spectral 
index of the dust thermal emission, and constrain the spatial distribution of dust grains 
with different sizes \citep[see, e.g.,][]{Perez12}. 
In addition, ALMA observations of optically thin molecular emission lines, such as, for example, 
the low-J rotational transitions of CO isotopologues, might constrain the density and kinematics of 
the gas in LkH$\alpha$~330 outer disk regions, and provide important information on the physical 
process that is causing the observed asymmetry.

\subsection{Azimuthal variations in the dust temperature}
  
Fluctuations in the dust temperature might cause azimuthal variations 
in the millimeter-wave dust emission even in the case in which the dust opacity and 
density are symmetric. 
In the case of a passive disk, i.e., a disk heated only by the central star, the temperature of 
the disk interior depends on the incident angle $\alpha_{inc}$ of the stellar radiation on the 
disk surface, so that larger values of $\alpha_{inc}$ lead to higher disk temperatures \cite[see, e.g.,][]{Chiang97}. 
Azimuthal fluctuations of the disk temperature can therefore happen if, 
for example, the disk is warped, or if one side of the disk is more flared than the other side, 
or if part of the outer disk lays in the shadow cast by some asymmetric structures in 
the inner disk. Such structures should have a scale of several AU, or 
tens of AU, to be able to affect the dust temperature across a large fraction 
of the disk far from the central star, and might therefore significantly differ from 
the small structures thought to be responsible of the short time-scale 
variability observed in  the disk mid-infrared emission \citep{Morales12}.

The presence of companions orbiting LkH$\alpha$~330 inside the millimeter cavity can 
affect the dust temperature in several ways. A warped disk might originate if the orbits of 
the companions are misaligned with the disk plane, as has been suggested to explain 
the warped geometry of the $\beta$ Pic disk \citep[see, e.g., ][]{Dawson11}. In addition, 
local variations of the disk flaring geometry might also occur at the inner and outer edges of 
the annular gaps  cleared by the companions \citep{Jang-Condell12}.
Finally,  the azimuthal variations in the disk surface density discussed in 
Section~\ref{sec:disc_inter} might also lead to variations in the dust temperature by 
changing the height of the disk surface. We investigated this latter point by performing 
radiative transfer calculations on the disk surface density provided by FARGO simulations
by adopting the "two-layer" model discussed in \cite{Isella09}.
We find that azimuthal fluctuations of a factor of 2 in the disk surface density might lead to 
variations in the disk mid plane temperature of about the same amplitude. 
However, the dependence of the disk temperature from the disk surface density is 
complex and needs to be investigated by adopting more accurate radiative transfer models. 
  
Finally, we note that asymmetries in the dust temperature might also occur if the disk is warped 
by a close stellar encounter \citep{Larwood01}, or, for example, if transient density 
fluctuations in the inner disk, e.g., the vortices that might originate from Rossby wave
instabilities at the outer edge of MRI dead-zone \citep{Lyra12}, cast a shadow on 
the outer disk, temporarily reducing the amount of stellar radiation that heats the 
disk mid plane. 

Asymmetries in the LkH$\alpha$~330 disk temperature can be measured 
by mapping the disk emission in optically thick molecular lines, such those 
corresponding to the low-J transitions of $^{12}$CO or CS. These observations 
are sensitive to the temperature of the disk layer at a depth of  $\tau \sim 1$, and, 
when combined with observations of optically thin lines,  would enable us to investigate whether 
the asymmetry observed in the millimeter-wave continuum emission toward LkH$\alpha$~330 is mainly  
caused by variations in the dust temperature, density, or opacity.

\section{Conclusions}
\label{sec:conc}

We present CARMA interferometric observations of the pre-main sequence star LkH$\alpha$~330
that reveal a lopsided ring in the 1.3~mm dust continuum emission. The ring has 
radius of about 100 AU and an azimuthal intensity variation of a factor of 2. 
By comparing the imaginary visibilities with parametric gaussian 
models, we find that the asymmetry in the dust emission traces a narrow circular 
arc, which extends in the azimuthal direction by about 90\arcdeg\ and accounts for about 
1/3 of the total disk flux at 1.3~mm.

Disk-planet interaction has been suggested as a possible mechanism to create 
large millimeter cavities, as that observed in the LkH$\alpha$~330 disk. 
We perform hydrodynamic simulations using FARGO  to investigate
whether this process might also explain the azimuthal asymmetry observed 
in the dust emission at 1.3~mm. We find that companions more massive than 
Jupiter orbiting within 70 AU from the central star might produce azimuthal 
asymmetries in the disk surface density between 70 AU and 90 AU, characterized by 
a density contrast similar to the variation in the dust intensity observed in 
LkH$\alpha$~330 disk.

We argue that the disk-planet interaction might also lead to azimuthal variations 
in the millimeter-wave dust opacity and dust temperature, so that the resulting 
dust continuum emission at 1.3~mm might depend on a complex 
interplay between asymmetries in the dust density, opacity, and temperature. 
Constraining the properties of possible unseen companions from millimeter-wave
observations therefore requires disentangling these three different contributions.
We suggest that this can be achieved by mapping the disk emission 
in both optically thin and thick molecular tracers, as well as in the dust continuum 
emission at multiple wavelengths between 1~mm and 1~cm.
 
We discussed alternative explanations for the observed asymmetry 
in LkH$\alpha$~330 which does not require the presence of companions: 
Rossby waves instabilities, baroclinic instabilities, disk warping, and 
disk shadowing. We argue that the first two processes might not apply to the 
outer regions of circumstellar disks, while the second two cannot be ruled out 
by existing observations. 

We conclude that, although the simulations of the interaction between the circumstellar 
material and possible companions orbiting within about 70 AU from LkH$\alpha$~330 
provide promising similarities with CARMA data, further observations in both the 
dust and molecular gas emission are required to derive firm conclusions on the origins 
of the asymmetry observed in LkH$\alpha$~330 disk.

\vspace{0.2cm}

\acknowledgments

We thank the OVRO/CARMA staff and the CARMA observers for their assistance 
in obtaining the data. Support for CARMA construction was derived from the Gordon and 
Betty Moore Foundation, the Kenneth T. and Eileen L. Norris Foundation, the James S. McDonnell 
Foundation, the Associates of the California Institute of Technology, the University of Chicago, 
the states of California, Illinois, and Maryland, and the National Science Foundation. Ongoing 
CARMA development and operations are supported by the National Science Foundation under 
a cooperative agreement, and by the CARMA partner universities.
We acknowledge support from the Owens Valley Radio Observatory, 
which is supported by the National Science Foundation through grant AST-1140063.
A.I. and J.M.C. acknowledge support from NSF award AST-1109334.
We thank Adam Kraus for sharing unpublished results.

\appendix

\section{Fourier transform of a 2-D Gaussian function}
We report here the derivation of the Fourier transform of a 2-D Gaussian function expressed in cartesian coordinates.
We start from the formulation of a 2-D Gaussian function centered in the origin and 
oriented along  the cartesian axes,
\begin{equation}
\label{eq:center_image}
g_0(x,y)= A \times  \exp \left( -  \frac{ x^2}{2\sigma_{maj}^2}\right)  \times \exp \left( - \frac{ y^2}{2\sigma_{min}^2}  \right), 
\end{equation}
where $A$ is the amplitude, and $\sigma_{maj}$ and $\sigma_{min}$ are 
the semi-major and semi-minor axes respectively.

We then derive the general formulation for a 2-D Gaussian function 
by applying a rotation by the angle $\theta$ followed by 
a translation by $x_0$ and $y_0$. In this way, the general expression for 
a 2-D Gaussian function assumes the form 
\begin{equation}
\label{eq:gen_image}
g(x,y)= A \times  \exp \left\{ - \frac{ [(x-x_0)\cos\theta+(y-y_0)\sin\theta]^2   }{2\sigma_{maj}^2} \right\}
	\times \exp \left\{ -\frac{[(x-x_0)\sin\theta-(y-y_0)\cos\theta]^2}{2\sigma_{min}^2} \right\}. 
\end{equation}

The Fourier transform of Equation~\ref{eq:gen_image} can be derived 
from the Fourier transform of Equation~\ref{eq:center_image} by applying the 
rotation and translation properties of the Fourier transform.
The first states that an anti-clockwise rotation of a function by an angle $\theta$ implies 
that its Fourier transform is also rotated anti-clockwise by the same angle. The second 
states that a shift in position of a function by an amount $x_0$  corresponds 
to a phase change in its Fourier transform by $\exp(i2\pi x_0 u)$.

The Fourier transform of Equation~\ref{eq:center_image} is
\begin{eqnarray}
G_0(u,v) & = & \int\int_{-\infty}^{+\infty} g_0(x,y) \exp[-i2\pi(ux+vy)] dx dy \\
	       & = & A \times \int\int_{-\infty}^{+\infty} \exp \left( - \frac{ x^2}{2\sigma_{maj}^2}\right) \exp(-i2\pi ux)dx  \times \exp \left( - \frac{ y^2}{2\sigma_{min}^2}  \right)  \exp(-i2\pi vy)dy\\
	       	& = & A \times F\left(\exp \left( - \frac{ x^2}{2\sigma_{maj}^2}\right) \right) \times F\left(\exp \left( - \frac{ y^2}{2\sigma_{min}^2}\right) \right),
\end{eqnarray}
 where $F\left(\exp \left( - \frac{ x^2}{2\sigma_{maj}^2}\right) \right) = \sqrt{2\pi} \sigma_{maj} \exp(-2\pi^2u^2 \sigma_{maj}^2)$ is the Fourier transform of a mono dimensional  Gaussian function. 
The Fourier transform of Equation~\ref{eq:center_image} therefore becomes
\begin{equation}
G_0(u,v) = A \times 2 \pi \sigma_{maj} \sigma_{min}  \exp{[-2\pi^2(u^2\sigma_{maj}^2 + v^2\sigma_{min}^2)]}
\end{equation}

We first apply the rotation defined by
\begin{equation}
	u' = u \cos\theta + v \sin\theta
\end{equation}
\begin{equation}
	v' = -u \sin\theta + v \cos\theta
\end{equation}
This leads to the equation
\begin{equation}
G(u,v) = A \times 2 \pi \sigma_{maj} \sigma_{min}  \exp{[-2\pi^2((u \cos\theta + v \sin\theta)^2\sigma_{maj}^2 + (-u \sin\theta + v \cos\theta)^2\sigma_{min}^2)].}
\end{equation}
The translation is then performed by applying a phase shift corresponding to $\exp[-2\pi i (x_0u+y_0 v)]$, which 
results in 
\begin{equation}
\label{eq:2df}
G(u,v) = A \times 2 \pi \sigma_{maj} \sigma_{min}  \exp{[-2\pi^2((u \cos\theta + v \sin\theta)^2\sigma_{maj}^2 + (-u \sin\theta + v \cos\theta)^2\sigma_{min}^2)]} \times \exp[-2\pi i (x_0u+y_0 v)]
\end{equation}

The real and imaginary part of Equation~\ref{eq:2df} can therefore be expressed as

\begin{equation}
\Re(G(u,v)) = A \times 2 \pi \sigma_{maj} \sigma_{min}  \exp{[-2\pi^2((u \cos\theta + v \sin\theta)^2\sigma_{maj}^2 + (-u \sin\theta + v \cos\theta)^2\sigma_{min}^2)]} \times \cos[-2\pi(x_0 u+y_o v)]
\end{equation}

\begin{equation}
\Im(G(u,v)) = A \times 2 \pi \sigma_{maj} \sigma_{min}  \exp{[-2\pi^2((u \cos\theta + v \sin\theta)^2\sigma_{maj}^2 + (-u \sin\theta + v \cos\theta)^2\sigma_{min}^2)]} \times \sin[-2\pi(x_0 u+y_o v)]
\end{equation}

\bibliographystyle{apj}
\bibliography{ref}

\end{document}